\documentclass[12pt]{iopart}
\usepackage{epsf}
\begin{document}

\title[K$^{*}(892)^{0}$
Production in Relativistic Heavy Ion Collisions at
$\sqrt{s_{NN}}$= 130 GeV]{K$^{*}(892)^{0}$ Production in
Relativistic Heavy Ion Collisions at $\sqrt{s_{NN}}$= 130 GeV}

\author{Patricia Fachini\dag\ for the STAR Collaboration\ddag}

\address{\dag\ Brookhaven National Laboratory, Bldg. 510A, Upton, NY 11973-5000,
USA}

\address{E-mail: pfachini@bnl.gov}

\address{\ddag\ For complete collaboration list see \cite{20}}

\begin{abstract}
Preliminary results on the K$^{*}(892)^{0} \rightarrow \pi$K
production using the mixed-event technique are presented. The
measurements are performed at mid-rapidity by the STAR detector in
$\sqrt{s_{NN}}$= 130 GeV Au-Au collisions at RHIC. The K$^{*0}$ to
negative hadron, kaon and $\phi$ ratios are obtained and compared
to the measurements in e$^{+}$e$^{-}$, pp and $\bar{\textrm{p}}$p
at various energies.
\end{abstract}



\maketitle

\section{Introduction}
The main motivation for studying heavy-ion collisions at high
energy is to investigate the properties of the strongly
interacting matter at high densities and temperatures. In
particular, the masses and widths of hadrons are expected to
change in hadronic or nuclear matter compared to their vacuum
values. Various theoretical models predict modification of hadron
masses and widths in a dense and hot medium \cite{28,1}. In this
context, the measurement of the properties of mesons whose
lifetimes are of the order of the lifetime of the dense matter may
be sensitive to the properties of the strongly interacting matter
in which they are produced. For example, model calculations show
that the K$^{*0}$/K ratio is sensitive to the mass modification of
particles in-medium and the dynamic evolution of the source
\cite{2}.

Due to the re-scattering of the daughter particles, the resonances
that decay into strongly interacting hadrons before thermal
freeze-out may not be reconstructed. However, resonances with
higher transverse momentum (p$_{\textrm{T}}$) may decay outside
the system. As a consequence, the measurement of the yields and
the p$_{\textrm{T}}$ distributions of resonances can provide
information on the time between chemical and kinetic freeze-out of
the system. In addition, we may be able to distinguish between a
sudden freeze-out \cite{3,4} or a smooth hadronic expansion
\cite{5,6} from a detailed comparison between the yield and the
p$_{\textrm{T}}$ distribution of resonances and stable hadrons.

On the other hand, due to the large population of $\pi$'s and K's
\cite{7,8,9} after chemical freeze-out, when the inelastic
interactions are too infrequent to change the particle species and
the total number of particles, the elastic interactions $\pi$K
$\rightarrow$ K$^{*0}$ $\rightarrow \pi$K increase the K$^{*0}$
population compensating for the K$^{*0}$ resonances that decay
before thermal freeze-out and may not be reconstructed.

Even though the hadronic decay modes for vector mesons are
dominant, only the leptonic decay modes have been studied
extensively mainly due to the large background from other produced
hadrons and the broad mass width. However, Monte Carlo
calculations have shown that the mixed-event technique
\cite{10,11} should allow a statistical measurement of both
K$^{*}(892)^{0}$ and $\overline{\textrm{K}}^{*}(892)^{0}$ at the
Relativistic Heavy Ion Collider (RHIC) energies because the
significance of the signal increases with the square root of the
number of events.

Preliminary results from the first measurement of such short-lived
resonance (c$\tau$ = 4 fm) via its hadronic decay channel in Au-Au
relativistic heavy-ion collisions at $\sqrt{s_{NN}}$= 130 GeV
using the STAR (Solenoidal Tracker At RHIC) detector at RHIC are
presented. The ratios K$^{*0}$/h$^{-}$, K$^{*0}$/K and
$\phi$/K$^{*0}$ are obtained and compared to the measurements in
e$^{+}$e$^{-}$, pp and $\bar{\textrm{p}}$p at various energies.

\section{Data Analysis}
The main STAR detector consists of a large Time Projection Chamber
(TPC) \cite{12} placed inside a uniform solenoidal magnetic field
that provides the measurement of charged particles. A
scintillating Central Trigger Barrel (CTB) that surrounds the TPC
is used as part of the centrality trigger by measuring the charged
particle multiplicity. Two hadronic calorimeters (ZDCs) located
upstream along the beam axis intercept spectator neutrons from the
collision and provide the minimum bias trigger. The
anti-correlation between the CTB and the ZDCs is used to trigger
on the event centrality.

In the summer of 2000, the first collisions between Au nuclei at
$\sqrt{s_{NN}}$= 130 GeV took place at RHIC. During this period,
the solenoidal magnet operated with a magnetic field of 0.25 T.
About 440K central and 230K minimum bias events that survive the
cuts mentioned below are used in this analysis. Central events
correspond to 14$\%$ of the hadronic cross-section. For this
analysis, only events with a vertex located within $\pm$95 cm of
the center of the TPC along the beam direction are selected, and
this assures uniform acceptance. The identification of the
daughter particles from the K$^{*0}$ decay is obtained by the
energy loss (dE/dx) in the gas of the TPC. The tracks are required
to cross the entire active area of the detector, to satisfy a
minimum number of points on the track cut, to point to the event
vertex within a certain accuracy and to have a transverse momentum
between 0.2 GeV/c and 2.0 GeV/c.

The decay channels K$^{*0} \rightarrow \pi^- \textrm{K}^+$ and
$\overline{\textrm{K}}^{*0} \rightarrow \pi^+ \textrm{K}^-$, each
with a branching ratio of 2/3, are selected for the measurements.
Due to limited statistics, the term K$^{*0}$ in this analysis
refers to the average of K$^{*0}$ and $\overline{\textrm{K}^{*0}}$
unless specified. The invariant mass of every opposite sign K$\pi$
pair is calculated for each event. The resulting invariant mass
distribution consists of the resonance signal and the
combinatorial background. The shape of the uncorrelated
combinatorial background is determined using the mixed-event
technique \cite{10,11}. Events are mixed for this background
calculation if their centrality triggers are similar and their
primary vertex locations are within 20 cm in the beam direction,
which minimizes possible fluctuations and distortions in the
uncorrelated background.

\section{Results}

\begin{figure}
\begin{center}
\epsfxsize=3.0in \epsfbox{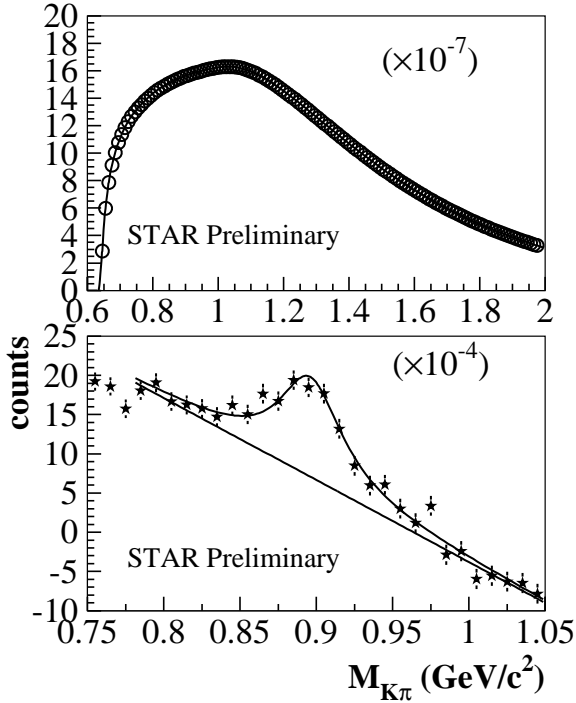}
\end{center}
\caption{\label{fig1}Top Panel: K$\pi$ invariant mass distribution
from same event pairs (open circles) and mixed-event pairs (solid
curve) for the 14$\%$ central collisions. The $x$-axis corresponds
to 10 MeV per bin and the data are scaled down by a factor of
$10^{-7}$. Bottom Panel: K$^{*0}$ invariant mass distribution
after background subtraction for the 14$\%$ central collisions.
The solid curve corresponds to a combination of a linear
background and a simple Breit-Wigner function are used to fit the
distribution with the resonance width and mass from the Particle
Data Book. The $x$-axis corresponds to 10 MeV per bin and the data
are scaled down by a factor of $10^{-4}$.}
\end{figure}

The invariant mass distributions from same-event K$\pi$ pairs and
from mixed-event pairs are shown in the upper panel of Fig
\ref{fig1} for the 14$\%$ most central events. There are more than
14 $\times$ 10$^9$ pairs of selected kaons and pions from these
central events. The signal to background ratio is about 1/1000 for
central and 1/200 for minimum bias events, which is significantly
lower than 1/4 for pp at $\sqrt{s_{{NN}}}$ = 63 GeV \cite{13}. The
lower panel of Fig \ref{fig1} depicts the K$\pi$ invariant mass
distribution after mixed-event background subtraction for the
14$\%$ most central hadronic interactions, where a 15 standard
deviation ($\sigma$) of the K$^{*0}$ signal above the background
fluctuation is observed.

The mixed-event technique \cite{10,11} describes the shape of the
uncorrelated background distribution; therefore, only the signal
and the correlated background are present in the distribution
after background subtraction. In the case of the K$\pi$ system,
besides the K$^{*}(892)^{0}$ resonance, there is the K$\pi$ and
S-wave correlation \cite{14}, which is not a resonance, and a long
list of higher resonant states. All these contribute to the K$\pi$
correlation. In addition, particle misidentification of the decay
products of $\rho$, $\omega$, $\eta$, $\eta$' and K$^0_S$ produce
correlations in the same-event distribution that are not present
in the mixed-event distribution used to estimate the background.
Studies using the HIJING model and an invariant mass distribution
produced from like-sign K$\pi$ pairs from the data are consistent
with the conclusion that the general features of the residual
combinatorial background in the K$\pi$ invariant mass distribution
after mixed-event background subtraction come from the
correlations mentioned. Since a quantitative description requires
the accurate knowledge of particle production and phase space
distributions that are not yet measured at RHIC energies, the
residual background was fit by both a linear and an exponential
function. Separately, the differences between the results obtained
from the two functions was about 20$\%$. These differences are
included in estimating the systematical uncertainties.

In order to obtain the m$_T$ - m$_0$ spectra, the K$^{*0}$ signal
is fit to a Breit-Wigner resonant function and corrected for
detector acceptance and efficiency. The acceptance and
reconstruction efficiency is determined by embedding simulated
tracks into real events at the raw data level, reconstructing the
full events and comparing the simulated input to the reconstructed
output. The acceptance and efficiency factor, $\epsilon$, depends
on the event centrality, the transverse momentum and the rapidity
of the parent and daughter particles. $\epsilon$ varies from under
15$\%$ for p$_T \simeq 0$ GeV/c to about 35$\%$ for p$_T \simeq
2.0$ GeV/c.

\begin{figure}
\begin{center}
\epsfxsize=3.0in \epsfbox{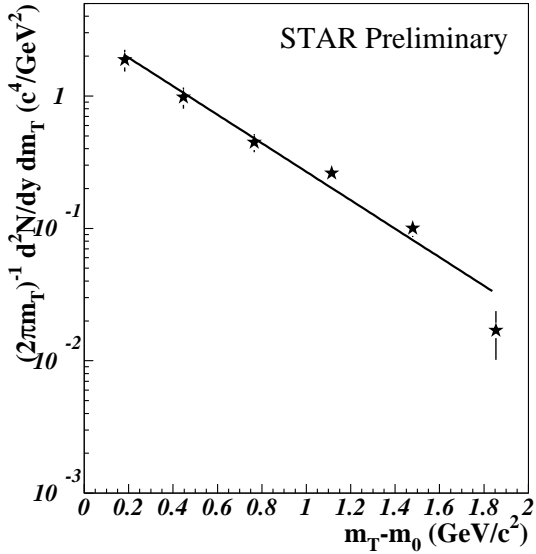}
\end{center}
\caption{\label{fig2} (K$^{*0} + \overline{\textrm{K}^{*0}})/2$
m$_T$ spectrum at mid-rapidity ($|y| <$ 0.5) for the 14$\%$ most
central interactions in the p$_T$ range from 0.4 GeV/c to 3.2
GeV/c. The errors shown are statistical only.}
\end{figure}

The d$^2$N/(2$\pi$m$_\textrm{T}$dm$_\textrm{T}$dy) distribution as
a function of m$_\textrm{T}$ - m$_0$ for central interactions at
mid-rapidity ($|y| <$ 0.5) is shown in Fig \ref{fig2}. An
exponential fit is used to extract the K$^{*0}$ yield per unit of
rapidity around mid-rapidity and the inverse slope (T). We obtain
dN/dy = 10.0 $\pm$ 0.8 (stat) and T = 0.41 $\pm$ 0.02 (stat) GeV.
The K$^{*0}$ and $\overline{\textrm{K}}^{*0}$ invariant mass
distributions integrated in p$_\textrm{T}$ are fit separately to a
Breit-Wigner resonant function and a linear residual background.
The ratio $\overline{\textrm{K}}^{*0}$/K$^{*0}$ = 0.92 $\pm$ 0.14
(stat) is obtained for central collisions. Therefore, the average
value between K$^{*0}$ and $\overline{\textrm{K}^{*0}}$ should
well represent the K$^{*}(892)^{0}$ production within our
statistics. The systematical uncertainty in dN/dy is estimated to
be 25$\%$ due to both detector effects and the uncertainty in the
background determination. By varying the analysis cuts and
studying the detector effects, the systematical uncertainty in the
inverse slope is estimated to be 10$\%$. Due to limited statistics
in the minimum bias data set, we assume that the K$^{*0}$
m$_\textrm{T}$ distribution for both central and minimum bias
events have the same shape, and in this way we can estimate the
K$^{*0}$ yield. The result is dN/dy = 4.5 $\pm$ 0.7 (stat) $\pm$
1.4 (sys). The difference between \cite{15} and the results
presented here is that we are able to measure the slope of 410
$\pm$ 20 (stat) MeV instead of an assumed 300 MeV slope, and we
use a linear function to describe the background instead of an
exponential function. Each of these lowers the yield by about
20$\%$.

The K$^{*0}$/h$^-$ ratios measured for central and minimum bias
events are compared to the measurements in pp \cite{13} and
e$^+$e$^-$ \cite{16,17,18}. The h$^-$ yield corresponds to the
corrected primary negatively charged hadrons at $|\eta| <$ 0.5
\cite{19}. At $\sqrt{s_{{NN}}}$ = 130 GeV, K$^{*0}$/h$^-$ = 0.042
$\pm$ 0.004 (stat) $\pm$ 0.01 (sys) for central collisions and
K$^{*0}$/h$^-$ = 0.059 $\pm$ 0.008 (stat) $\pm$ 0.019 (sys) for
minimum bias events. These results are compared to K$^{*0}$/$\pi$
= 0.057 $\pm$ 0.012 measured in pp at $\sqrt{s_{{NN}}}$ = 63 GeV
\cite{13} and K$^{*0}$/$\pi$ = 0.044 $\pm$ 0.003 measured in
e$^+$e$^-$ at $\sqrt{s_{{NN}}}$ = 91 GeV \cite{16,17,18}. The
error in the measurements in pp and e$^+$e$^-$ corresponds to the
quadratic sum of the systematical and statistical errors. Noting
that about 80$\%$ of h$^-$ are pions at RHIC energies, we observe
that the K$^{*0}$/h$^-$ ratio is comparable to the K$^{*0}$/$\pi$
ratios measured in pp and e$^+$e$^-$. We do not observe the
enhancement in the K$^{*0}$ production observed for other strange
particles as function of increasing energies and colliding systems
\cite{20}.

\begin{figure}
\begin{center}
\epsfxsize=3.1in \epsfbox{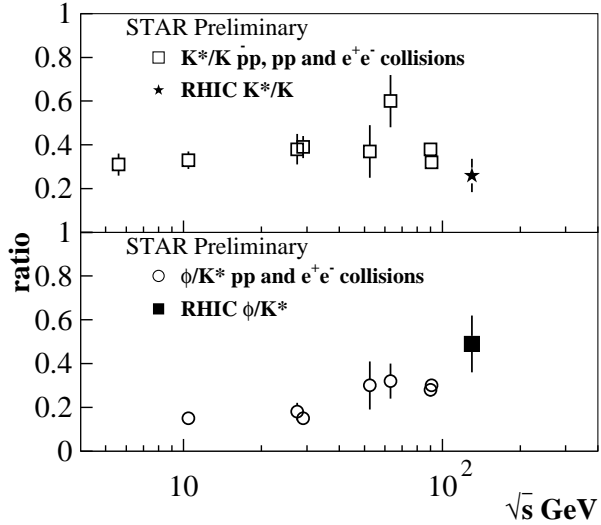}
\end{center}
\caption{\label{fig3} K$^{*0}$/K and $\phi$/K$^{*0}$ ratios
measured in different colliding systems at various energies. The
ratios are from measurements in e$^+$e$^-$ collisions at 10.45 GeV
\cite{21}, 29 GeV \cite{22}, 90 GeV \cite{16} and 91 GeV
\cite{17,18} beam energies, $\bar{\textrm{p}}$p at 5.6 GeV
\cite{23} and pp at 27.5 GeV \cite{24}, 52.5 GeV \cite{25} and 63
GeV \cite{13}. The errors shown correspond to the quadratic sum of
the statistical and systematical errors.}
\end{figure}

Fig \ref{fig3} shows the K$^{*0}$/K and $\phi$/K$^{*0}$ ratios
measured in different colliding systems at various energies. The
K$^{*0}$/K ratio is interesting because K$^{*0}$ and K have
similar quark content and differ mainly in mass and spin. Fig
\ref{fig3} shows that the ratio K$^{*0}$/K = 0.26 $\pm$ 0.03
(stat) $\pm$ 0.07 (sys) measured in central Au-Au collisions at
130 GeV beam energy is compatible or even lower than the
measurements in $\bar{\textrm{p}}$p, pp and e$^+$e$^-$ at lower
energies.

The $\phi$/K$^{*0}$ ratio measures the strangeness suppression in
near ideal conditions since $\Delta$S = 1 with hidden strangeness
in the $\phi$, and there is only a small mass difference. Fig
\ref{fig3} shows an increase of the ratio $\phi$/K$^{*0}$ = 0.49
$\pm$ 0.05 (stat) $\pm$ 0.12 (sys) measured in central Au-Au
collisions at 130 GeV beam energy \cite{26} compared to the
measurements in pp and e$^+$e$^-$ at lower energies. However, this
increase may not be solely related to strangeness suppression due
to additional effects on short lived resonances in heavy-ion
collisions. It is also interesting to note that the
$\phi$/K$^{*0}$ ratio already has a rising tendency as a function
of beam energy in pp and e$^+$e$^-$.

Since the K$^{*0}$ lifetime is short (c$\tau$ = 4 fm) and
comparable to the time scale of the evolution of the system in
relativistic heavy-ion collisions, the K$^{*0}$ survival
probability needs to be taken into account. This survival
probability depends on the expansion time between chemical and
thermal freeze-out, the source size and the K$^{*0}$ transverse
momentum. Should the K$^{*0}$ decay between chemical and thermal
freeze-out, the daughters from the decay may re-scatter and the
K$^{*0}$ will not be reconstructed. On the other hand, chemical
freeze-out is only truly defined for non-resonant long-lived
particles. Elastic interactions such as $\pi$K $\rightarrow$
K$^{*0}$ $\rightarrow \pi$K are still effective and regenerate the
K$^{*0}$ until thermal freeze-out. Due to these additional effects
on short lived resonances in heavy-ion collisions, the K$^{*0}$
measurement at RHIC may provide information on the expansion time
between chemical and thermal freeze-out. However, final state
interactions of the hadronic decay products destroy the early-time
information carried by these decays. Assuming that the difference
in the K$^{*0}$/K ratio measured in Au-Au collisions at RHIC and
the ratios measured in pp and e$^+$e$^-$ is due to the K$^{*0}$
survival probability, our measurement is consistent with an
expansion time of only a few fm between chemical and thermal
freeze-out (sudden freeze-out). In the scenario of a long
expansion time between chemical and thermal freeze-out ($\sim$20
fm \cite{5,6}) and without K$^{*0}$ regeneration, not only is the
measured K$^{*0}$ production reduced but there is also a low
p$_\textrm{T}$ depression. This results in a large effective
inverse slope than would otherwise be expected. However, the
measured K$^{*0}$ inverse slope is similar to the $\phi$ inverse
slope \cite{26} and higher than that of kaons \cite{20}. Hence,
our measurement at RHIC is consistent with either a sudden
freeze-out with no K$^{*0}$ regeneration or a long expansion time
scenario with significant K$^{*0}$ regeneration. In the
statistical model, the measured K$^{*0}$ should be consistent with
the condition at kinetic freeze-out instead of chemical
freeze-out. The fact that the statistical model \cite{27}
reproduces our measurement is consistent with a short expansion
time between chemical and thermal freeze-out, since the
statistical model provides particle yields at chemical freeze-out.

\section{Conclusions}
Preliminary results on K$^{*}(892)^{0}$ and
$\overline{\textrm{K}}^{*}(892)^{0}$ production measured at
mid-rapidity by the STAR detector in $\sqrt{s_{NN}}$= 130 GeV
Au-Au collisions at RHIC were presented. Despite the short
lifetime (c$\tau$ = 4 fm), the data show significant K$^{*0}$
production. However, we do not observe an enhancement in the
K$^{*0}$ production as observed for other strange particles for
increasing energies and differing colliding systems. If the
difference in the K$^{*0}$/K ratio measured in Au-Au collisions at
RHIC and the ratios measured in pp and e$^+$e$^-$ is due to the
K$^{*0}$ survival probability, our measurement is consistent with
a sudden freeze-out. Hence, the K$^{*0}$ production at RHIC rules
out a long expansion time between chemical and thermal freeze-out
unless there is K$^{*0}$ regeneration, due to elastic interactions
after chemical freeze-out.

Finally, the study of resonances like the K$^{*0}$ may provide
important information on the collision dynamics. The improvement
of the uncertainties in the K$^{*0}$ and $\phi$ measurements and
the measurements of other resonances with different properties may
help in understanding the development of the system after chemical
freeze-out.

\section*{Acknowledgments}
We wish to thank the RHIC Operations Group
and the RHIC Computing Facility at Brookhaven National Laboratory,
and the National Energy Research Scientific Computing Center at
Lawrence Berkeley Laboratory for their support. This work was
supported by the Division of Nuclear Physics and the Division of
High Energy Physics of the Office of Science of the U.S.Department
of Energy, the United States National Science Foundation, the
Bundesministerium fuer Bildung und Forschung of Germany, the
Institut National de la Physique Nucleaire et de la Physique des
Particules of France, the United Kingdom Engineering and Physical
Sciences Research Council, Funda\c c\~ao de Amparo a Pesquisa do
Estado de S\~ao Paulo, Brazil, and the Russian Ministry of Science
and Technology.

\end{document}